\documentclass[preprint,twocolumn]{aastex6}
\newcommand{\tar}{K2-155}
\usepackage{amsmath,amssymb}
\usepackage{graphicx} 
\usepackage{natbib} 
\usepackage{aas_macros} 
\slugcomment{}
\shortauthors{Hirano et al.}
\shorttitle{K2-155}
\begin{document}
\title{K2-155: A Bright Metal-Poor M Dwarf with Three Transiting Super-Earths}
\author{
Teruyuki Hirano\altaffilmark{1},
Fei Dai\altaffilmark{2,3},
John H. Livingston\altaffilmark{4},
Yuka Fujii\altaffilmark{5},
William D. Cochran\altaffilmark{6},
Michael Endl\altaffilmark{6},
Davide Gandolfi\altaffilmark{7},
Seth Redfield\altaffilmark{8},
Joshua N.\ Winn\altaffilmark{3},
Eike W. Guenther\altaffilmark{9},
Jorge Prieto-Arranz\altaffilmark{10,11}, 
Simon Albrecht\altaffilmark{12}, 
Oscar Barragan\altaffilmark{7},
Juan Cabrera\altaffilmark{13},
P.\ Wilson Cauley\altaffilmark{14},
Szilard Csizmadia\altaffilmark{13},
Hans Deeg\altaffilmark{10,11},
Philipp Eigm\"uller\altaffilmark{13},
Anders Erikson\altaffilmark{13},
Malcolm Fridlund\altaffilmark{15,16},
Akihiko Fukui\altaffilmark{17},
Sascha Grziwa\altaffilmark{18},
Artie P. Hatzes\altaffilmark{9},
Judith Korth\altaffilmark{18},
Norio Narita\altaffilmark{4,19,20},
David Nespral\altaffilmark{10,11},
Prajwal Niraula\altaffilmark{8},
Grzegorz Nowak\altaffilmark{10,11},
Martin P\"atzold\altaffilmark{18},
Enric Palle\altaffilmark{10,11},
Carina M. Persson\altaffilmark{16},
Heike Rauer\altaffilmark{13,21},
Ignasi Ribas\altaffilmark{22},
Alexis M. S. Smith\altaffilmark{13},
Vincent Van Eylen\altaffilmark{15}
}
\altaffiltext{1}{Department of Earth and Planetary Sciences, Tokyo Institute of Technology,
2-12-1 Ookayama, Meguro-ku, Tokyo 152-8551, Japan}
\email{hirano@geo.titech.ac.jp}
\altaffiltext{2}{Department of Physics, and Kavli Institute for Astrophysics and Space Research, Massachusetts Institute of Technology, Cambridge, MA 02139, USA}
\altaffiltext{3}{Department of Astrophysical Sciences, Princeton University, 4 Ivy Lane, Princeton, NJ 08544, USA}
\altaffiltext{4}{Department of Astronomy, Graduate School of Science, The University of Tokyo, Hongo 7-3-1, Bunkyo-ku, Tokyo, 113-0033, Japan}
\altaffiltext{5}{Earth-Life Science Institute, Tokyo Institute of Technology, Tokyo, 152-8550, Japan}
\altaffiltext{6}{Department of Astronomy and McDonald Observatory, University of Texas at Austin, 2515 Speedway,~Stop~C1400, Austin, TX 78712, USA}
\altaffiltext{7}{Dipartimento di Fisica, Universit\`a di Torino, via P. Giuria 1, 10125 Torino, Italy}
\altaffiltext{8}{Astronomy Department and Van Vleck Observatory, Wesleyan University, Middletown, CT 06459, USA}
\altaffiltext{9}{Th\"uringer Landessternwarte Tautenburg, Sternwarte 5, D-07778 Tautenberg, Germany}
\altaffiltext{10}{Instituto de Astrof\'\i sica de Canarias, C/\,V\'\i a L\'actea s/n, 38205 La Laguna, Spain}
\altaffiltext{11}{Departamento de Astrof\'isica, Universidad de La Laguna, 38206 La Laguna, Spain}

\altaffiltext{12}{Stellar Astrophysics Centre, Department of Physics and Astronomy, Aarhus University, Ny Munkegade 120, DK-8000 Aarhus C, Denmark}
\altaffiltext{13}{Institute of Planetary Research, German Aerospace Center, Rutherfordstrasse 2, 12489 Berlin, Germany}
\altaffiltext{14}{School of Earth and Space Exploration, Arizona State University, Tempe, AZ 85281, USA}
\altaffiltext{15}{Leiden Observatory, Leiden University, 2333CA Leiden, The Netherlands}
\altaffiltext{16}{Department of Space, Earth and Environment, Chalmers University of Technology, Onsala Space Observatory, 439 92 Onsala, Sweden}
\altaffiltext{17}{Okayama Astrophysical Observatory, National Astronomical Observatory of Japan, Asakuchi, Okayama 719-0232, Japan}
\altaffiltext{18}{Rheinisches Institut f\"ur Umweltforschung an der Universit\"at zu K\"oln, Aachener Strasse 209, 50931 K\"oln, Germany}
\altaffiltext{19}{Astrobiology Center, NINS, 2-21-1 Osawa, Mitaka, Tokyo 181-8588, Japan}
\altaffiltext{20}{National Astronomical Observatory of Japan, NINS, 2-21-1 Osawa, Mitaka, Tokyo 181-8588, Japan}
\altaffiltext{21}{Center for Astronomy and Astrophysics, TU Berlin, Hardenbergstr. 36, 10623 Berlin, Germany}
\altaffiltext{22}{Institut de Ci\`{e}ncies de l'Espai (CSIC-IEEC), Carrer de Can Magrans, Campus UAB, 08193 Bellaterra, Spain}

\begin{abstract}
We report on the discovery of three transiting super-Earths around
\tar\ (EPIC 210897587), a relatively bright early M dwarf ($V=12.81$ mag) observed
during Campaign 13 of the NASA {\it K2} mission. To characterize the
system and validate the planet candidates, we conducted speckle
imaging and high-dispersion optical spectroscopy, including radial velocity measurements.  
Based on the {\it K2} light curve and the spectroscopic characterization of the host
star, the planet sizes and orbital periods are
$1.55_{-0.17}^{+0.20}\,R_\oplus$ and $6.34365\pm 0.00028$ days for the
inner planet; $1.95_{-0.22}^{+0.27}\,R_\oplus$ and $13.85402\pm 0.00088$ 
days for the middle planet; and
$1.64_{-0.17}^{+0.18}\,R_\oplus$ and $40.6835\pm 0.0031$ days for the
outer planet.  The outer planet (\tar d) is near the habitable zone,
with an insolation $1.67\pm 0.38$ times that of the Earth.  The
planet's radius falls within the range between that of smaller rocky
planets and larger gas-rich planets.  To assess the habitability of
this planet, we present a series of 3D global climate simulations
assuming that \tar d is tidally locked and has an Earth-like composition and 
atmosphere. We find that the planet can maintain a moderate surface temperature 
if the insolation proves to be smaller than $\sim 1.5$ times that of the Earth.  
Doppler mass measurements, transit spectroscopy, and other
follow-up observations should be rewarding, since \tar\ is one of the
optically brightest M dwarfs known to harbor transiting planets.
\end{abstract}

\section{Introduction\label{s:intro}}\label{s:intro}

Nearby stars are always attractive targets for the characterization of
exoplanets of all sizes.  Nearby M dwarfs are especially attractive
because their small sizes lead to larger transit and Doppler signals,
and because the habitable zone occurs at relatively short orbital
periods.  However, the number of optically bright M dwarfs known to
have transiting planets is still small.  
As of January 2018, there are only a handful of transiting planets orbiting
M dwarfs that are bright enough for further follow-up observations 
\citep[e.g., $V\lesssim 14$ mag;][]{2004ApJ...617..580B,2012A&A...546A..27B,2015ApJ...804...10C,2015Natur.527..204B}.

After the failure of two reaction wheels, the {\it Kepler} spacecraft
ended its original mission and was repurposed to conduct another
transit survey known as the {\it K2} mission
\citep{2014PASP..126..398H}.  This new survey is examining a series of
star fields around the ecliptic.  Together these fields cover a much
wider area of the sky than the original mission, but each field is
observed for a shorter duration (about 80 days) than the original
mission (4 years).  Because of the wider sky coverage, it has been
possible to observe a larger sample of bright and nearby stars.  This
has led to many new planet discoveries, including planets around
low-mass stars \citep[see, e.g.,][]{2015ApJ...804...10C,
  2015ApJ...809...25M, 2016ApJ...820...41H, 2017AJ....154..207D}.

KESPRINT is one of several large collaborations that are detecting
planet candidates using {\it K2} data and performing follow-up
observations to validate the candidates and measure planet masses
\citep[see, e.g.,][]{2017A&A...604A..16F, 2017AJ....154..123G,
2017A&A...608A..93G, 2017arXiv171007203L}.  This latest KESPRINT
paper focuses on \tar\ (EPIC 210897587), a bright M dwarf ($V=12.81$) observed during
Campaign 13 of the {\it K2} mission.  Table~\ref{hyo1} draws together
the basic parameters of the star from the literature
\citep{2017AJ....153..166Z, 2016yCat.2336....0H, 2006AJ....131.1163S,
  2012yCat.2311....0C}.  The {\it K2} data reveal that \tar\ is a
candidate host of three transiting super-Earths.  Systems with
multiple planetary candidates are known to have a very low probability
of being false positives \citep[FPs;][]{2012ApJ...750..112L}.  The
follow-up observations presented in this paper confirm that the
planets are very likely to be genuine.

We organize this paper as follows. Section \ref{s:k2} describes the
reduction of the {\it K2} data and detection of the three planet
candidates.  Section \ref{s:obs} presents follow-up observations using
ground-based telescopes, including high-resolution speckle imaging and
high-dispersion optical spectroscopy.  Section \ref{s:analysis}
presents our best estimates of the stellar and planetary parameters
based on all the data. Section \ref{s:discussion} compares \tar\ with
another recently discovered planetary system, K2-3, and discusses the
potential habitability of the outer planet as well as the prospects
for future follow-up observations. Section~\ref{s:summary} summarizes
all our findings.

\section{Light Curve Extraction and Transit Search\label{s:k2}}\label{s:k2}

\begin{figure*}
\centering
\includegraphics[width=18.5cm]{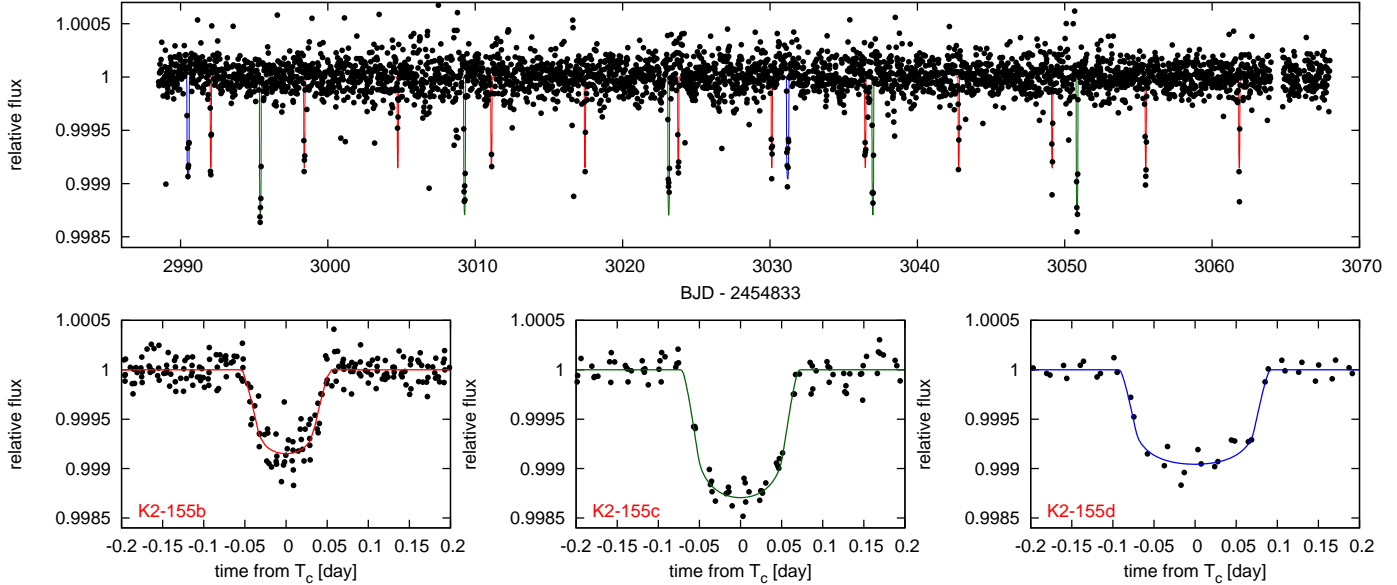}
\caption{
{\it Upper:} Normalized light curve of \tar\ obtained during {\it K2} campaign 13.
The vertical lines indicate the times of planetary transits.
{\it Lower:} Folded light curve for each planet.
}
\label{fig:lightcurve}
\end{figure*}

\tar\ was observed in the long cadence mode in {\it K2} Campaign 13
from UT 2017 March 8 to 2017 May 27.  Our light curve extraction
and transit search pipeline were described in detail by 
\citet{2017AJ....154..226D}
and Livingston et al. (under review).
In short, we used the observed motion of the center-of-light on the
detector to detrend the systematic flux variation introduced by the
rolling motion of the spacecraft, similar to
\citet{2014PASP..126..948V}.  We searched the detrended light curve
(the upper panel of Figure \ref{fig:lightcurve}) for periodic transit
signals with the Box-Least-Squares (BLS) algorithm
\citep{2002A&A...391..369K}.  We found three transiting planet
candidates after iteratively searching for the strongest peak in the
BLS periodogram and removing the signal of the detected planets.  We
then scrutinized the light curve and did not see odd-even variations
or secondary eclipses which would be produced by FPs such as a blended
eclipsing binary or a hierarchical eclipsing binary.

\section{Observations\label{s:obs}}\label{s:obs}

\subsection{Speckle Observations\label{s:wiyn}}\label{s:wiyn}

\begin{figure}
\centering
\includegraphics[width=8.5cm]{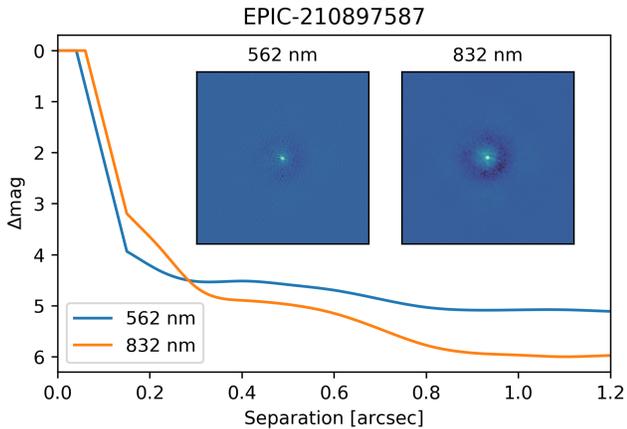}
\caption{
5-$\sigma$ contrast curves of the reconstructed images for \tar\ (insets),
based on speckle observations with WIYN/NESSI.
}
\label{fig:contrast}
\end{figure}

We performed high-resolution imaging on the night of
UT~2017~September~5 with the WIYN~3.5m telescope and the NASA
Exoplanet Star and Speckle Imager (NESSI; Scott et al., in
preparation). This instrument uses high-speed electron-multiplying
CCDs (EMCCDs) to obtain 40~ms exposures simultaneously in two bands: a
`blue' band centered at 562 nm with a width of 44 nm, and a `red' band
centered at 832 nm with a width of 40 nm.  The pixel scales of the
'blue' and `red' EMCCDs are $0\farcs0175649$ pix$^{-1}$ and
$0\farcs0181887$ pix$^{-1}$, respectively.  We observed \tar\ along
with nearby point-source calibrator stars, spaced closely in time.
Following the procedures described by \citet{2011AJ....142...19H}, we
used the calibrator images to compute reconstructed
$256\,\mathrm{pix}\,\times\,256\,\mathrm{pix}$ images in each band,
corresponding to $4\farcs6\times4\farcs6$.

No additional light sources were detected in the reconstructed images
of \tar.  We measured the background sensitivity of the reconstructed
images using a series of concentric annuli centered on the target
star, resulting in 5-$\sigma$ sensitivity limits (in delta-magnitudes)
as a function of angular separation. The 5-$\sigma$ contrast curve as
well as the reconstructed image in each band are displayed in Figure
\ref{fig:contrast}.

\subsection{High Dispersion Spectroscopy\label{s:hires}}\label{s:hires}

We observed \tar\ with the Tull Coude Spectrograph
\citep{1995PASP..107..251T} on the McDonald Observatory 2.7~m Harlan
J. Smith Telescope on UT 2017 September 14 and 2017 October 14.  The
spectrograph is a cross-dispersed echelle instrument covering
375-1020~nm, with increasingly larger inter-order gaps longward of
570~nm.  A 1.2 arc-second wide slit projects to 2 pixels on the CCD
detector, resulting in a spectral resolving power of 60,000.  On each
date, three successive short exposures were obtained in order to
reject cosmic ray events.  We used an exposure meter to obtain an
accurate flux-weighted barycentric correction, and to establish an
exposure time resulting in a signal-to-noise ratio (SNR) of about 30
per pixel.  Bracketing exposures of a Th-Ar hollow cathode lamp were
obtained in order to generate a wavelength calibration and to remove
spectrograph drifts.  The raw data were processed using IRAF routines
to remove the bias level, inter-order scattered light, and
pixel-to-pixel (``flat field") CCD sensitivity variations.  We traced
the apertures for each spectral order and used an optimal-extraction
algorithm to obtain the detected stellar flux as a function of
wavelength.

We obtained four high-resolution spectra with the FIbre-fed
Echelle Spectrograph 
\citep[FIES;][]{1999anot.conf...71F,2014AN....335...41T}
on the 2.56 m Nordic Optical Telescope (NOT) at the Observatorio del Roque de los Muchachos, 
La Palma (Spain). The observations were carried out on UT 2017 December 24, 25, 27,
and 2018 January 10 as part of the observing programs 2017B/059 (OPTICON) and 56-209 (CAT). 
We used the $1\farcs3$ high-resolution fiber ($\lambda/\Delta                                            
\lambda=67,000$) and set the exposure time to three times 20 minutes, following
the same observing strategy as in \citet{2015A&A...576A..11G}.
We traced the RV drift of the instrument by acquiring Th-Ar spectra immediately
before and after each observation. The data were reduced using
standard IRAF and IDL routines. The SNR of the extracted spectra is
about 20 per pixel at 5500\,\AA. 

\section{Data Analysis\label{s:analysis}}\label{s:analysis}

\subsection{Stellar Parameters\label{s:parameter}}\label{s:parameter}

\begin{table}[tb]
\begin{center}
\caption{Stellar Parameters of \tar}\label{hyo1}
\begin{tabular}{lcc}
\hline\hline
Parameter & Value & Source \\\hline
\multicolumn{2}{l}{\it (Identifiers)} & \\
EPIC & 210897587 & \\
2MASS & J04215245+2121131 & \\
\hline
\multicolumn{2}{l}{\it (Stellar Parameters from the Literature)} & \\
RA (J2000) & 04:21:52.49 & UCAC5\\
Dec (J2000) & +21:21:12.95 & UCAC5 \\
$\mu_\alpha\cos\delta$ (mas yr$^{-1}$) & $199.6\pm1.3$ & UCAC5 \\
$\mu_\delta$ (mas yr$^{-1}$) & $-77.3\pm1.3$ & UCAC5 \\
$B$ (mag) & $14.073\pm0.051$ & APASS \\
$V$ (mag) & $12.806\pm0.046$ & APASS \\
$g^\prime$ (mag) & $13.491 \pm 0.047$ & APASS  \\
$r^\prime$ (mag) & $12.286 \pm 0.059$ & APASS  \\
$J$ (mag) & $10.274\pm0.024$ & 2MASS \\
$H$ (mag) & $9.686\pm0.022$ & 2MASS \\
$K_\mathrm{s}$ (mag) & $9.496\pm0.017$ & 2MASS \\
$W1$ (mag) & $9.435\pm0.023$ & WISE \\
$W2$ (mag) & $9.404\pm0.019$ & WISE \\
$W3$ (mag) & $9.343\pm0.041$ & WISE \\
$W4$ (mag) & $>8.317$ & WISE \\
\hline
\multicolumn{2}{l}{\it (Spectroscopic and Derived Parameters)} &\\
$T_{\rm eff}$ (K) & $3919\pm70$ & This work\\
$[\mathrm{Fe/H}]$ (dex) & $-0.42\pm0.12$ & This work\\
$R_\star$ ($R_\odot$) & $0.526\pm0.053$ & This work\\
$M_\star$ ($M_\odot$) & $0.540\pm0.056$ & This work\\
$\log g$ (cgs) & $4.732\pm0.046$ & This work\\
$\rho_\star$ ($\rho_\odot$) & $3.84\pm0.79$ & This work\\
$L_\star$ ($L_\odot$) & $0.059\pm0.013$ & This work\\
distance (pc) & $62.3\pm9.3$ & This work\\
RV (km s$^{-1}$) & $19.34\pm0.16$ & This work\\
$U$ (km s$^{-1}$) & $24.6\pm2.2$ & This work\\
$V$ (km s$^{-1}$) & $-39.9\pm8.2$ & This work\\
$W$ (km s$^{-1}$) & $27.5\pm4.1$ & This work\\
\hline
\end{tabular}
\end{center}
\end{table}

We analyzed the high resolution spectra taken by McDonald 2.7~m/Tull
and estimated the stellar parameters. Following
\citet{2017arXiv171003239H}, we used {\tt SpecMatch-Emp}
\citep{2017ApJ...836...77Y} to derive the spectroscopic parameters for
\tar. {\tt SpecMatch-Emp} tries to match the input observed spectrum
to hundreds of library spectra covering a wide range of stellar
parameters, and finds a subset of stellar spectra that best match the
input spectrum. The stellar parameters (the effective temperature
$T_\mathrm{eff}$, radius $R_\star$, and metallicity [Fe/H]) are then
estimated by interpolating the parameters for the best-matched spectra. 
We analyzed each of the two Tull spectra separately with
{\tt SpecMatch-Emp}, finding that the results were consistent with
each other to within 1-$\sigma$\footnote{The library spectra in 
{\tt SpecMatch-Emp} were secured by Keck/HIRES. As we discussed in 
\citet{2017arXiv171003239H}, we checked the validity of 
{\tt SpecMatch-Emp} for the Tull spectra by putting several Tull spectra 
(mainly K dwarfs) into the code, and found that the output parameters are 
all consistent with the parameters estimated by the {\tt Kea} code
\citep{2016PASP..128i4502E} within 2-$\sigma$. }. 
To check for the accuracy of our analysis, we also applied the same 
technique to the FIES spectrum, and obtained a fully consistent result.

To derive the stellar mass $M_\star$, surface gravity $\log g$,
density $\rho_\star$, and luminosity $L_\star$, we used a Monte Carlo
technique based on the empirical relations for the stellar parameters
of M dwarfs derived by \citet{2015ApJ...804...64M}. Assuming that
$T_\mathrm{eff}$, $R_\star$, and [Fe/H] returned by {\tt SpecMatch-Emp} 
follow independent Gaussian distributions, we
perturbed those parameters to estimate $M_\star$, $\log g$,
$\rho_\star$, and $L_\star$ through the absolute $K_s$-band magnitude,
which we estimated as $5.52\pm0.33$ mag.  The result is shown in Table
\ref{hyo1}, which also includes the distance of \tar, estimated from
the absolute and apparent $K_s$-band magnitudes.

\begin{figure}
\centering
\includegraphics[width=8.5cm]{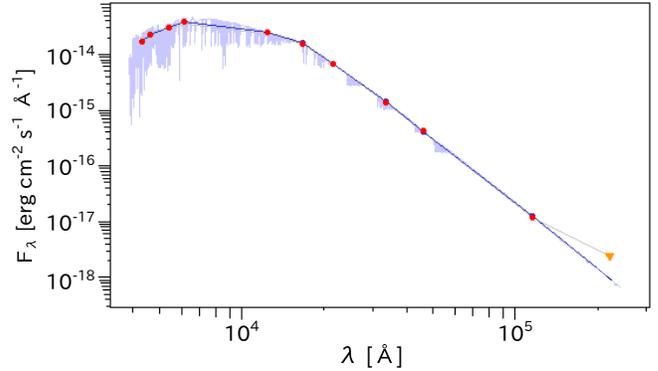}
\caption{
Spectral energy distribution of \tar. The fluxes based on the magnitudes listed in Table \ref{hyo1} 
are plotted by the red points. The best-fitting BT-SETTL CIFIST synthetic spectrum is shown 
in grey. The WISE flux at 22 micron (orange triangle) is an upper limit and is not used for the fit.
}
\label{fig:sed}
\end{figure}
Following the method described in \citet{Gandolfi2008}, we derived the interstellar reddening along 
the line of sight ($A_\mathrm{v}$) and obtained an independent estimate of $T_\mathrm{eff}$
and $\log g$ for \tar. Briefly, we built the spectral energy distribution (SED) of the star using 
the APASS $B, g^\prime, V, r^\prime$, 2MASS $J, H, K_\mathrm{s}$, and WISE $W1, W2, W3$ 
magnitudes listed in Table \ref{hyo1}. 
We retrieved the Johnson $BV$, Sloan $g^\prime$ and $r^\prime$, 2MASS $JHK_s$, WISE $W1$, $W2$, 
$W3$, and $W4$ transmission curves, and absolute flux calibration constants from the Asiago 
Database on Photometric Systems \citep{2000A&AS..147..361M, 2002Ap&SS.280...77F} 
and from \citet{2010AJ....140.1868W}. 
We simultaneously fitted the SED for $T_\mathrm{eff}$ and $A_\mathrm{v}$ using the BT-SETTL CIFIST 
synthetic spectra from \citet{Baraffe2015}. 
We assumed a total-to-selective extinction of 3.1 (normal 
interstellar extinction) and adopted the reddening law from \citet{Cardelli1989}. We found a reddening 
of $A_\mathrm{v}$\,=\,0.095$\,\pm$\,0.050\,mag, effective temperature of 
$T_\mathrm{eff}$\,=\,4200\,$\pm$\,200\,K, and surface gravity of $\log g=5.5\pm1.0$ (cgs). 
The result of the SED fit is shown in Figure \ref{fig:sed}. 
Both $T_\mathrm{eff}$ and $\log g$ are in good agreement with the spectroscopically derived 
values, corroborating our results.
Note that $A_\mathrm{v}$ is explored in the positive range, and thus its estimate could be 
biased towards higher values.

\subsection{RV Measurements and Star's Membership\label{s:cc}}\label{s:cc}

In order to estimate the absolute radial velocities (RVs) of the star and
check for any secondary lines in the high resolution spectra, 
we cross-correlated the Tull spectra against the M2 numerical
mask \citep[e.g.,][]{2013A&A...549A.109B}, developed for the precise RV measurement
for HARPS-like spectrographs. To take into account the possible wavelength drift
of the spectrograph within the night, we also cross-correlated the spectral segment
including strong telluric absorptions ($6860-6930\,\mathrm{\AA}$) against a
theoretical telluric template created by the line-by-line radiative transfer model
\citep[LBLRTM;][]{2005JQSRT..91..233C}. The absolute RV of \tar\ was calculated by
subtracting the telluric RV value (whose magnitude is $\sim 0.5$ km~s$^{-1}$) from
the stellar RV value, both of which were estimated
by inspecting the peaks of the cross-correlation functions (CCFs).

\begin{table}[tb]
\begin{center}
\small
\caption{Results of RV Measurements}\label{tab:rv}
\begin{tabular}{lcccc}
\hline\hline
BJD$_\mathrm{TDB}$ & RV & RV error & RV Type & Instrument \\
($-2450000.0$) & (km s$^{-1}$) & (km s$^{-1}$) & & \\\hline
8010.907364 & $19.416$ & 0.274 & absolute & Tull \\
8040.879045 & $19.305$ & 0.201 & absolute & Tull \\
8112.545268 & $0.000$ & 0.027 & relative & FIES \\
8113.547023 & $-0.046$ & 0.031 & relative & FIES \\
8115.497963 & $-0.011$ & 0.022 & relative & FIES \\
8129.443988 & $-0.028$ & 0.022 & relative & FIES \\
\hline
\end{tabular}
\end{center}
\end{table}

Table \ref{tab:rv} lists the absolute RVs measured from the Tull spectra. 
The mean absolute RV ($19.34\pm0.16$ km s$^{-1}$) by Tull is consistent with the value 
reported in the literature within 2-$\sigma$ \citep[$20.3\pm0.5$ km s$^{-1}$;][]{2007AN....328..889K}, 
which also suggests that there has been no significant RV variation of the star
over the course of $\sim 10$ years. 

For the FIES spectra, we measured relative RVs using multi-order cross-correlations. 
In doing so, we 
first derived the RVs by cross-correlating the spectra against the first spectrum. 
We then applied the RV shift and co-added the individual spectra to obtain the combined spectrum. 
Finally, the co-added spectrum is used to extract the final RVs.
Thus derived relative RVs are listed in Table \ref{tab:rv}.

\begin{figure}
\centering
\includegraphics[width=8.5cm]{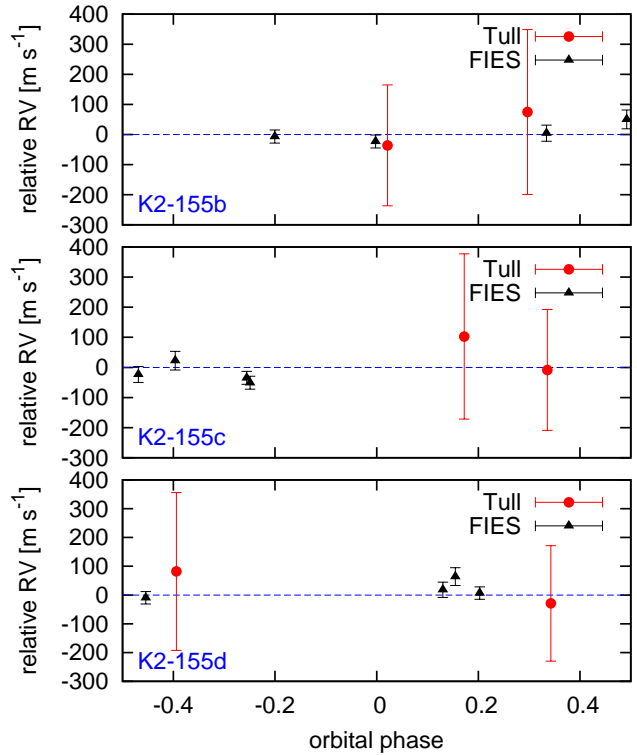}
\caption{
Relative RVs measured by Tull (red circles) and FIES (black triangle), folded by 
the orbital periods of inner (top), middle (middle), and outer 
(bottom) planets, respectively. 
}
\label{fig:rv}
\end{figure}
To place an upper limit on the mass of any companion, we estimated the upper
limit of the RV semi-amplitude $K$ by fitting the data folded 
by the orbital periods of the planet candidates. In the fit, we introduced an RV offset
parameter for each of the two datasets. 
This yielded $K=-8 \pm 19$ m~s$^{-1}$, $K=-38_{-42}^{+39}$ m~s$^{-1}$, and 
$K=-26_{-25}^{+23}$ m~s$^{-1}$ for the inner, middle, and outer planet candidates, 
respectively, indicating that the observed RVs are consistent with $K=0$ m~s$^{-1}$
within $\sim 1$-$\sigma$ for all three periods. 
Thus the eclipsing binary (EB) scenario for the three planet candidates
is strongly constrained. 
The 2-$\sigma$ upper limits on $K$ translate to the upper limits on the companion's 
mass of $60\,M_\oplus$, $102\,M_\oplus$, and $75\,M_\oplus$, respectively, 
all of which fall on the planetary regime. 
Since the RV data should be fitted for the three companions simultaneously, 
however, these cannot be interpreted as the mass upper limits of the 
planet candidates.

The absence of secondary lines in the CCF for Tull spectra also allows us to place an
upper limit on the brightness of any close-orbiting companions.  To do
so, we fitted the observed CCFs by two components: (1) the observed
CCF after flattening the continuum to its average value, and (2) the
scaled and Doppler-shifted version of the same CCF to mimic a possible
faint companion. Here, we implicitly assume that the spectrum of the
hypothetical companion is similar to that of the primary (i.e., a
late-type star).  For a given Doppler shift for the secondary line
(relative RV $>15$ km~s$^{-1}$), we computed the possible
contamination of a secondary peak, and looked for the maximum
contamination flux from a hypothetical companion.  We conclude that
the contamination is no more than $2\,\%$ of the primary star's flux
in the visible band, which corresponds to lowest mass stars ($\sim
0.1\,M_\odot$) for the case of \tar.  This is a good constraint on the
presence of close-in companion(s), but when the companion has a long
orbital distance, the relative RV between the primary and secondary
stars becomes small (relative RV $\lesssim 10$ km s$^{-1}$), and we
are not capable of constraining its flux by the present analysis.

The coordinates of \tar\ place it near the same line of sight as the
Hyades open cluster.  However, \tar\ does not share the same
metallicity, proper motion, or radial velocity as typical Hyades
stars.  The metallicity and mean proper motion of the Hyades are
reported to be $\mathrm{[Fe/H]}=0.14\pm0.05$
\citep{1998A&A...331...81P} and $\mu_\alpha\cos\delta=1.4\pm3.7$ mas
yr$^{-1}$ and $\mu_\delta=-4.3\pm4.4$ mas yr$^{-1}$
\citep{2014A&A...564A..79D}, respectively.  Together with the absolute
RV\footnote{
The averaged absolute RV of Hyades members are reported to be $39.29
\pm 0.25$ km~s$^{-1}$ \citep{2002A&A...389..871D}. }, we conclude that
\tar\ is in the background of the Hyades.

Based on the coordinates, proper motion, distance, and RV of \tar, we
also computed the galactic space velocity $(U, V, W)$ to the Local
Standard of Rest (LSR) as in Table \ref{hyo1}.  The space velocity
components are in agreement with those of both thick disk and thin
disk stellar populations \citep[e.g.,][]{2004AN....325....3F}, making
it impossible to tell on this basis to which population \tar\ belongs.
The low stellar metallicity is more consistent with the thick disk.

\subsection{Planetary Parameters\label{s:lightcurve}}\label{s:lightcurve}

To determine the planetary parameters, we compared two available light
curves: our own light curve as produced in Section \ref{s:k2} and the
publicly available light curve provided by
\citet{2014PASP..126..948V}.  The two light curves have almost the
same noise level, although our light curve exhibits a slightly larger
scatter at the beginning of the {\it K2} observation.  We decided to
adopt the light curve of \citet{2014PASP..126..948V} for subsequent
analysis.

The fitting procedure of the K2 light curve was described in detail by
\citet{2015ApJ...799....9H}, which we summarize here.  We first split
the light curve into chunks each spanning approximately 5 days and
fitted each chunk after removing transit signals with a fifth-order
polynomial to detrend and obtain the normalized light curve. Then,
based on the preliminary ephemerides obtained in Section \ref{s:k2},
we extracted small segments of the normalized light curve, which cover
the transits of each planet candidate as well as the flux baselines on
both sides spanning 2.0 times the transit durations.

For each planet candidate, we simultaneously fitted all the segments
to estimate the global parameters common to all the segments as well
as segment-specific parameters.  The global parameters are the scaled
semi-major axis $a/R_\star$, transit impact parameter $b$,
limb-darkening parameters for the quadratic law ($u_1+u_2$ and
$u_1-u_2$), 
orbital eccentricity and argument of periastron ($e\cos\omega$ and $e\sin \omega$), 
and planet-to-star radius ratio $R_p/R_\star$. To take into
account possible transit timing variations (TTVs), we allowed the
mid-transit time $T_c$ to float freely for each light curve
segment. We also introduced additional parameters describing the
baseline flux variation for each segment, which we assumed to be a
linear function of time. 

The goodness of fit was assessed with the $\chi^2$ statistic:
\begin{eqnarray}
\label{eq:chisq}
\chi^2 &=& \sum_{i}\frac{(f_{\mathrm{obs},i}-f_{\mathrm{calc},i})^2}{\sigma_i^2},
\end{eqnarray}
where $f_{\mathrm{obs},i}$ and $f_{\mathrm{calc},i}$ are the observed and calculated flux,
and $\sigma_i$ is the flux uncertainty.
For the transit model, we integrated the analytic light curve model of \citet{2009ApJ...690....1O}
over the 30-minute averaging interval of {\it K2} observations.
We sampled the posterior distributions of the parameters using our implementation of the Markov
Chain Monte Carlo (MCMC) technique \citep{2015ApJ...799....9H}.
In the code, all the free parameters are first optimized simultaneously by
Powell's conjugate direction method \citep[e.g.,][]{1992nrca.book.....P}, and the flux
baseline parameters are held fixed at the best-fitting values.
We then took $10^6$ MCMC steps for each planet candidate with all the
parameters being allowed to adjust.
We imposed prior distributions for $u_1+u_2$ and $u_1-u_2$ adopted from
the table by \citet{2013A&A...552A..16C}, assuming Gaussian functions with
widths of 0.20. 
Since close-in planets in multi-planet systems are known to have low 
eccentricities \citep{2015ApJ...808..126V}, we also imposed Gaussian priors 
on $e\cos\omega$ and $e\sin \omega$ with their centers and widths being 0 
and 0.05, respectively. 
For the other parameters, we assumed uniform priors. 
The reported parameter values and $\pm 1\sigma$ errors are based on the 50, 15.87
and 84.13 percentile levels of the marginalized posterior distributions.
Table \ref{hyo2} gives the results.

\begin{figure}
\centering
\includegraphics[width=8.5cm]{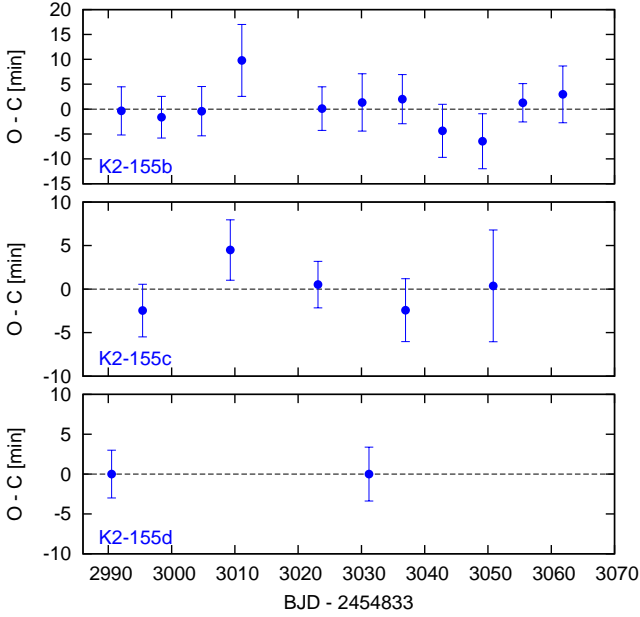}
\caption{
$O-C$ diagrams for the mid-transit times $T_c$. There is no evidence of
significant TTVs.
}
\label{fig:ttv}
\end{figure}

Based on the mid-transit times, we calculated the ephemerides ($P$ and
$T_{c,0}$) for each planet under the assumption of a constant period.
Figure \ref{fig:ttv} shows the observed minus calculated ($O-C$) $T_c$
plots for the three candidates.  The period ratio between \tar b and
\tar c is somewhat close to $1:2$, but Figure \ref{fig:ttv} exhibits
no clear sign of TTVs. In the bottom panels of Figure
\ref{fig:lightcurve}, we display the folded transits along with the
model light curves (solid lines) based on the parameters given in
Table \ref{hyo2}.

\begin{table}[tb]
\caption{Fitting and Planetary Parameters}\label{hyo2}
\centering
\begin{tabular}{lc}
\hline\hline
Parameter & Value \\\hline
\multicolumn{2}{l}{\bf \tar b}  \\
$P$ (days) & $6.34365\pm 0.00028$  \\
$T_{c,0}$ ($\mathrm{BJD}-2454833$) & $2985.7153\pm 0.0021$ \\
$a/R_\star$ & $20.3_{-6.1}^{+3.2}$ \\
$b$ & $0.50_{-0.33}^{+0.30}$ \\
$R_p/R_\star$ & $0.0271_{-0.0012}^{+0.0023}$ \\
$u_1$ & $0.36\pm 0.13$ \\
$u_2$ & $0.41\pm 0.14$ \\
$R_p$ ($R_\oplus$) & $1.55_{-0.17}^{+0.20}$ \\
$a$ (AU) & $0.0546\pm 0.0019$ \\
$S_p$ ($S_\oplus$) & $19.9\pm 4.5$ \\\hline
\multicolumn{2}{l}{\bf \tar c}  \\
$P$ (days) & $13.85402\pm 0.00088$ \\
$T_{c,0}$ ($\mathrm{BJD}-2454833$) & $2981.5643\pm 0.0025$ \\
$a/R_\star$ & $30.0_{-9.8}^{+6.2}$ \\
$b$ & $0.57_{-0.39}^{+0.27}$ \\
$R_p/R_\star$ & $0.0339_{-0.0016}^{+0.0031}$  \\
$u_1$ & $0.33\pm 0.13$ \\
$u_2$ & $0.40_{-0.14}^{+0.13}$ \\
$R_p$ ($R_\oplus$) & $1.95_{-0.22}^{+0.27}$ \\
$a$ (AU) & $0.0920\pm 0.0032$ \\
$S_p$ ($S_\oplus$) & $7.0\pm 1.6$ \\\hline
\multicolumn{2}{l}{\bf \tar d} \\
$P$ (days) & $40.6835\pm 0.0031$  \\
$T_{c,0}$ ($\mathrm{BJD}-2454833$) & $2949.8324\pm 0.0048$ \\
$a/R_\star$ & $73.5_{-15.9}^{+8.4}$  \\
$b$ & $0.41_{-0.28}^{+0.30}$ \\
$R_p/R_\star$ & $0.0286_{-0.0010}^{+0.0015}$  \\
$u_1$ & $0.32\pm0.12$ \\
$u_2$ & $0.38_{-0.13}^{+0.14}$ \\
$R_p$ ($R_\oplus$) & $1.64_{-0.17}^{+0.18}$ \\
$a$ (AU) & $0.1886\pm 0.0066$ \\
$S_p$ ($S_\oplus$) & $1.67\pm 0.38$ \\
\hline
\end{tabular}
\end{table}

\subsection{Validating Planets\label{s:validation}}\label{validation}

Since \tar\ has three planet candidates, the probability that any of
the candidates will turn out to be a FP is extremely low.
\citet{2012ApJ...750..112L} calculated the odds that the systems of
multiple transiting planet candidates are FPs.  For three-planet
systems, they found that fewer than one such system is expected to
contain a FP among the entire {\it Kepler} sample.  In this sense the
presence of three candidates is self-validating.  Below, we
investigate the constraints on FP scenarios based on direct
follow-up observations rather than the statistical argument of
\citet{2012ApJ...750..112L}.

\begin{figure*}
\centering
\includegraphics[width=13cm]{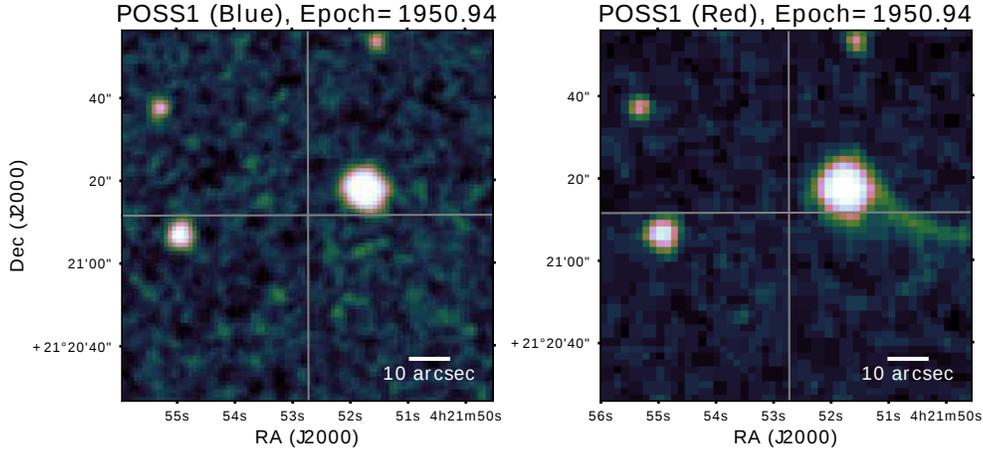}
\caption{The POSS1 Blue and Red images of \tar\ obtained in 1950. North is up and East is to the left. 
The gray lines indicate the position of the target at $\mathrm{epoch}=2017.0$.
The brightest star on the right is \tar. We note that 5-$\sigma$ contrasts between \tar\ and
the region at the current position of \tar\ were $5-5.5$ mag. 
}
\label{fig:poss1}
\end{figure*}
As shown in Section \ref{s:cc}, the absence of a large RV variation ($\gtrsim 100$ m~s$^{-1}$) 
as well as a secondary peak in the CCFs implied that the transit signals are not caused by 
a stellar companion orbiting and occulting \tar\ (i.e., EB). 
The remaining possible FP scenarios are background eclipsing binaries (BEB)
and hierarchical-triple eclipsing binaries (HEB).
However, these scenarios are also constrained by the lack of bright nearby sources
in the reconstructed image from the speckle observations (Figure \ref{fig:contrast}).
In addition, checking the POSS1 archival images taken in 1950 (Figure \ref{fig:poss1}), 
we found no bright star at the current position of \tar, verifying that no background sources
are hidden in the reconstructed image of \tar\ by chance alignment.

Since the speckle observations with WIYN/NESSI are only able to find companions in the proximity
of the target, it is still possible that a fainter object at a large separation is blended
in the {\it K2} aperture, which could be responsible for the transit-like photometric signals.
We thus searched for fainter objects within $20^{\prime\prime}$ from \tar\ using
the SDSS photometric catalog \citep{2015ApJS..219...12A}. As a consequence,
we identified five stars within $12^{\prime\prime}-20^{\prime\prime}$ from \tar,
but all of those stars have $r-$band magnitudes (similar to the {\it Kepler} magnitudes)
fainter than $20$ mag. 
The $r-$magnitude of \tar\ is $12.437\pm0.002$ mag, and thus the maximum magnitude
that can produce an eclipse depth of $0.1\%$ is $r=19.9$ mag ($100\%$ occultation). 
Therefore, we conclude that \tar\ is the source of the transit signals.

Regarding the HEB scenario, the speckle observations achieved a 5-$\sigma$
contrast of 4.2 mag (562 nm) at $0\farcs2$, corresponding to the mass
upper limit of $\approx 0.1\,M_\odot$ for a possible bound companion 
\citep[e.g.,][]{2008ApJS..178...89D} at the projected
separation of $\approx 12$ AU and further. There is still a possibility, however, that
a very late-type star is orbiting \tar\ at an orbital distance of $1-12$ AU;
for instance, a $0.1\,M_\odot$ star with $P=2$ yr exerts an RV semi-amplitude
of only $\approx 3$ km~s$^{-1}$, which could be overlooked in the RV data 
(Table \ref{tab:rv}). But even if this is the case and the bound
later-type star is responsible for the transit signals, the depths of these
candidates correspond to those of ``planets".

The fact that \tar\ is transited by the three planet candidates is corroborated by comparing
the mean stellar density inferred from spectroscopy ($\rho_\star=3.84\pm0.79\,\rho_\odot$) with
the mean stellar density implied by the transit modeling.
The scaled semi-major axes $a/R_\star$ in Table \ref{hyo2} are translated into the mean
stellar densities of $2.8_{-1.8}^{+1.6}\,\rho_\odot$ for the inner, $1.9_{-1.3}^{+1.4}\,\rho_\odot$ 
for the middle, and $3.2_{-1.7}^{+1.2}\,\rho_\odot$ for the outer planet, respectively.
Hence, the stellar densities estimated from transit modelings are consistent
with the spectroscopic density for \tar\ within about 1-$\sigma$, but would be inconsistent
with later-type stars; according the observed mass-radius relation for M dwarfs
\citep[e.g.,][]{2015ApJ...804...64M}, the mean density of mid-to-late M dwarfs with
$M_\star<0.4\,M_\odot$ is higher than $\approx 6\,\rho_\odot$.

To quantify the false positive probability (FPP) of each planet candidate, we used the statistical
framework implemented in the {\tt vespa} software package \citep{2012ApJ...761....6M, 2015ascl.soft03011M}.
This code simulates FP scenarios using the {\tt TRILEGAL} Galaxy model
\citep{2005A&A...436..895G} and assesses the likelihoods of EB, BEB, and HEB scenarios.
The inputs to {\tt vespa} are the phase-folded light curve, the size of the photometric aperture,
contrast curves from high resolution imaging, the maximum secondary eclipse depth allowed by the
{\it K2} light curve, as well as the broadband photometry and spectroscopic stellar parameters of
the host star. The FPPs computed by {\tt vespa} are below $10^{-5}$ for all three
planet candidates of \tar. However, because {\tt vespa} considers each planet individually,
it does not take into account the ``multiplicity boost'' suggested by \citet{2012ApJ...750..112L},
who found that planet candidates belonging to stars with 3 or more candidates are {\it a priori}
$\sim$100 times more likely to be valid planets than single candidates.
This means that the FPPs computed by {\tt vespa} are likely to be overestimated
by two orders of magnitude. Each of \tar's three planet candidates are therefore below the
fiducial validation threshold of 1\% FPP by some 5 orders of magnitude. Thus, all three candidates
are quantitatively validated, in addition to our independent determination of the low likelihoods of
FP scenarios. We conclude that the three candidates of \tar\ are indeed {\it bona fide} planets.


\section{Discussion\label{s:discussion}}\label{s:discussion}

\subsection{Comparison with the K2-3 System}

\begin{table*}[tb]
\centering
\caption{Comparison between the \tar\ System and K2-3 System}\label{hyo3}
\begin{tabular}{lccc|lccc}
\hline\hline
Planet  & $P$ (days) & $R_p$ ($R_\oplus$) & $S_p$ ($S_\oplus$) & Planet & $P$ (days) & $R_p$ ($R_\oplus$) & $S_p$ ($S_\oplus$) \\\hline
\tar b & $6.34365\pm 0.00028$ & $1.55_{-0.17}^{+0.20}$ & $19.9\pm 4.5$ & K2-3b & $10.05403_{-0.00025}^{+0.00026}$ & $1.90\pm0.20$ & $8.7\pm2.0$ \\
\tar c & $13.85402\pm 0.00088$ & $1.95_{-0.22}^{+0.27}$ & $7.0\pm 1.6$ & K2-3c & $24.6454\pm0.0013$ & $1.52\pm0.17$ & $2.64\pm0.59$ \\
\tar d & $40.6835\pm 0.0031$ & $1.64_{-0.17}^{+0.18}$ & $1.67\pm 0.38$ & K2-3d & $44.55612\pm0.00021$ & $1.35\pm0.16$ & $1.20\pm0.27$ \\
\hline
\end{tabular}
\end{table*}

\tar\ is similar to K2-3 \citep{2015ApJ...804...10C} in many aspects. Both stars are
relatively bright early M dwarfs ($V=12.81$ mag for \tar\ and $V=12.17$ mag for K2-3) hosting three
transiting super-Earths. Table \ref{hyo3} summarizes the planetary parameters for the two systems
\citep{2016ApJ...823..115D, 2017arXiv171003239H, 2016AJ....152..171F}.
In both systems, the outermost transiting planets receive stellar insolations that are slightly
higher than the solar insolation on the Earth ($1\, S_\oplus$), but less than $\approx 2\, S_\oplus$.
One difference between the two systems is the size ordering of the planets.
For \tar\ the middle planet is the largest, while for K2-3 the inner planet is the largest.

\subsection{Habitability of \tar d}
The outer planet (\tar d) has a relatively long orbital period and receives a stellar
insolation flux similar to that of Earth ($S=1.67\pm 0.38\, S_\oplus$).
This implies that \tar d is located in or near the habitable zone around \tar.
Another factor affecting potential habitability is whether the planet has a solid
surface or is smothered by a massive atmosphere.
\citet{2015ApJ...801...41R} noted that planets larger than $1.6\,R_\oplus$ are
likely to possess volatile-rich atmospheres.  The size of \tar d ($\approx 1.64\, R_\oplus$) is very
close to this boundary.  It also falls within the observed ``valley" in the planet radius distribution \citep{2017AJ....154..109F},
making it a particularly interesting target for characterizing its internal structure and atmosphere.
Recently, \citet{2017arXiv171005398V} confirmed the presence of the radius gap with 
more precise measurements of stellar and thus planetary radii, and found that its dependence
on the orbital period suggests that it is likely caused by photoevaporation.

At this point, it is unclear whether \tar d is rocky or not, until we
make a precise mass measurement by RV or TTV observations.  We decided
to investigate whether the planet would be habitable if it does turn
out to have an Earth-like composition and atmosphere.  Three-dimensional
(3D) global climate simulations have shown that tidally-locked planets
can have a moderate surface temperature in a wide range of orbital
distance due to the climate-stabilizing effects of dayside clouds
\citep{2013ApJ...771L..45Y, 2016ApJ...819...84K}, but may undergo the
classical moist greenhouse state at the higher end of the incident
flux.
Recent studies suggested that for an Earth-sized planet with a nitrogen-dominated atmosphere around an
M1 star, this occurs when the total incident flux exceeds $\approx 1.4\,S_\oplus$
\citep{2017ApJ...848..100F, 2017ApJ...845....5K}.

\begin{figure}
\centering
\includegraphics[width=8.5cm]{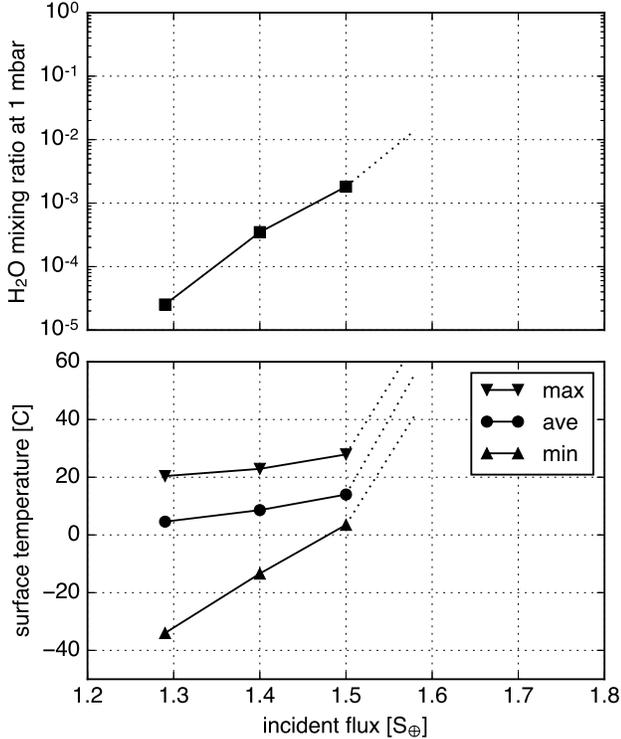}
\caption{
Results of 3D global climate simulations for \tar d. We plot the water mixing ratio at 1 mbar (upper)
and global maximum, average, and minimum surface temperatures (lower) as a function of insolation 
flux on \tar d.
When the insolation exceeds $\approx 1.5\,S_\oplus$, 
both surface temperature and water mixing ratio continue to increase until they eventually enter 
the regime where the model is invalid due to high temperature and high humidity. 
}
\label{fig:GCM}
\end{figure}

In the earlier studies, the planet was assumed to be an Earth-sized one ($1.0\,R_\oplus$)
with the Earth's surface gravity.
In order to find a possible climate specifically for \tar d, 
we ran a series of global climate simulations using a 3D General Circulation Model (GCM) 
ROCKE-3D \citep{2017ApJS..231...12W}, 
fixing the planetary parameters at the values of \tar d.
The setup is equivalent to the model coupled with a dynamic ocean (900 meter depth) used in 
\citet{2017ApJ...848..100F}, except that the planetary size and the rotation/orbital period are 
specified for \tar d.
Namely, the planetary radius is fixed at $1.6\,R_{\oplus }$ and its mass is set to $4.2\,M_{\oplus}$
based on the empirical relation of \citet{2014ApJ...783L...6W}.
Given the proximity to the star, the three planets around \tar\ are expected to be tidally locked
\citep[e.g.,][]{1993Icar..101..108K, 2017CeMDA.129..509B}, and thus
the rotation period is assumed to be equal to the orbital period (40.6835 days).
For the input stellar spectrum, we adopted the PHOENIX atmosphere model
\citep[BT-SETTL;][]{2013MSAIS..24..128A} for which we adopted the stellar parameters of \tar.
We assumed the planet is covered with a thermodynamic ocean, and assumed a 1 bar N$_2$ atmosphere 
and 1 ppm of CO$_2$ as in \citet{2017ApJ...848..100F}. 
We increased the incident flux from $1.29\,S_{\oplus}$ (the 1-$\sigma$ lower limit of $S_p$ in Table \ref{hyo2})
to $1.67\,S_{\oplus}$ (the best fit value) and checked the range that allows for the planet 
to have a moderate surface temperature.
The upper panel of Figure \ref{fig:GCM} shows the water mixing ratio at 1 mbar for varying incident flux,
while the lower panel presents the corresponding maximum, global average, and minimum surface temperatures.

Similarly to previous works for an Earth analog, equilibrium climates were secured
up to $S_p\approx 1.5\,S_{\oplus}$. 
Above this limit, the model surface temperature continues to increase until it
enters the regime where the model is invalid and the simulation stops.
When the insolation is close to or lower than $1.5\,S_{\oplus}$, the surface temperature remains moderate,
comparable to that of the Earth.
The upper humidity increases gradually as incident flux increases, and is about to cross the classical moist
greenhouse state at about $1.5\,S_{\oplus}$.
Thus, \tar d has a potential of being habitable if the incident flux turns out to be
close to the lower end within the uncertainty range, though the actual habitability also depends
on other factors including its atmosphere, water content, and initial stellar luminosity 
\citep[e.g.,][]{2015AsBio..15..119L, 2015NatGe...8..177T}.
Another important factor that potentially affects the habitability is the presence of frequent flares 
of the host star \citep[e.g.,][]{2017ApJ...841..124V}. We inspected the {\it K2} light curve
of \tar, but found no such event over the course of 80 days.



\subsection{Prospects on Future Follow-up Observations}
Given the brightness for an M dwarf, \tar\ is an attractive target for
future follow-up studies, including Doppler mass measurements and
transit photometry.  Among M dwarfs ($T_\mathrm{eff}\leq 4000$ K) with
transiting planets, \tar\ is the fourth brightest star in the
$V$-band, after GJ\,436, K2-3, and GJ\,3470.  It is also the second
brightest M dwarf in optical passbands (after K2-3) having a possibly
habitable transiting planet ($S_p\lesssim2.0\,S_\oplus$).  

Based on the empirical mass estimates by \citet{2014ApJ...783L...6W}, we
estimate the RV semi-amplitudes of the planets to be $K\approx 2.1$
m~s$^{-1}$, $2.0$ m~s$^{-1}$, and $1.2$ m~s$^{-1}$ for the inner,
middle, and outer planets, respectively, 
suggesting that the masses of at least inner two planets could be constrained 
by gathering a large number ($\sim 50-100$) data points with a precision of
$2-3$ m~s$^{-1}$ \citep[e.g.,][]{2017A&A...608A..93G}. 
Although challenging, observations of M dwarfs of similar magnitude (e.g., K2-3)
have shown that RV precisions of $2-3$ m~s$^{-1}$ were achieved
by TNG/HARPS-N and Magellan/PSF \citep[e.g.,][]{2015A&A...581L...7A, 2016ApJ...823..115D},
and thus these measurements seem feasible with high precision spectrographs
on $8-10$ m telescopes such as Keck/HIRES. 
Considering that the three planets
straddle the rocky to volatile-rich boundary
\citep{2015ApJ...801...41R} and also the radius gap suggested by
\citet{2017AJ....154..109F}, the comparison between the mean densities
of these planets may provide some insight into the origin of close-in
super-Earths in multi-planet systems.

\begin{figure*}
\centering
\includegraphics[width=16cm]{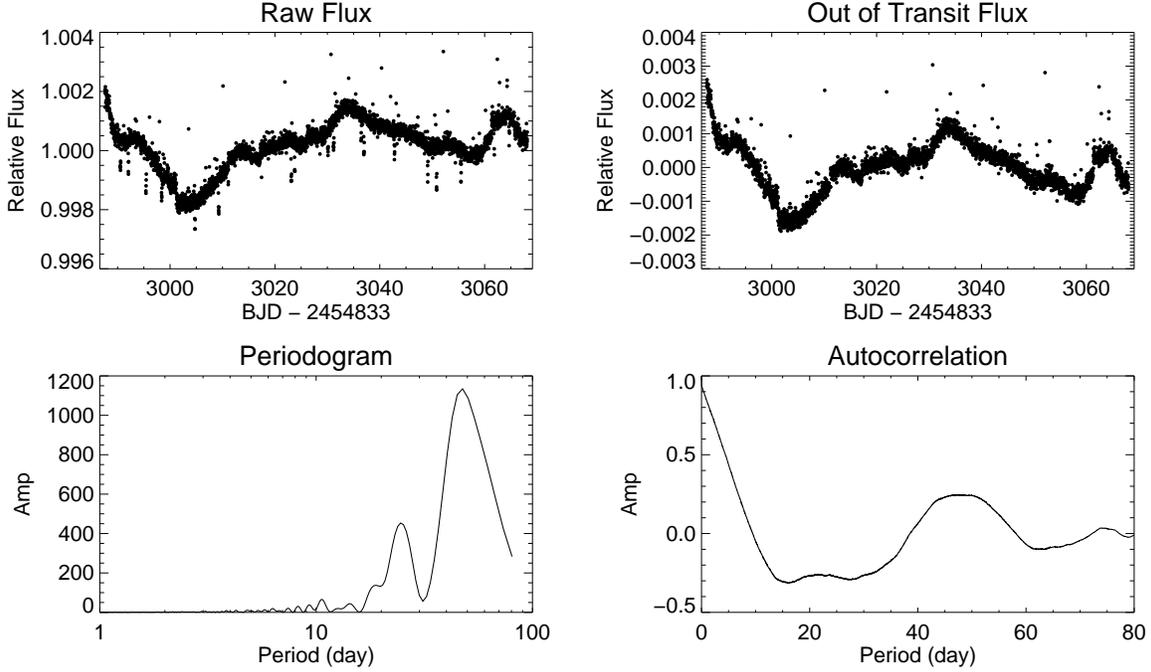}
\caption{\tar's light curve before detrending (top two panels), and its 
Lomb-Scargle periodogram (bottom left) and auto-correlation function (bottom right). 
}
\label{fig:rotation}
\end{figure*}
We note that the expected RV jitter for \tar\ is small. 
In order to estimate the rotation period of the star, we computed
the Lomb-Scargle periodogram and auto-correlation function of the 
light curve \citep{2014ApJS..211...24M}, both of which are shown in Figure \ref{fig:rotation}. 
Both methods yielded similar estimates for the rotation period 
($P_\mathrm{rot}=47.5_{-9.0}^{+19.3}$ days and 
$P_\mathrm{rot}=46.2_{-5.6}^{+8.1}$ days, respectively), although
the detected period could be an alias given the short 
observing span of {\it K2} ($\sim 80$ days). 
Using this tentative rotation period together with
the stellar radius ($R_\star=0.526\,R_\odot$), we
estimate the equatorial velocity of the star as $\approx 0.58$ km s$^{-1}$, 
which is the maximum value for the projected stellar spin
velocity ($v\sin i$). The {\it K2} light curve exhibits the
photometric variation amplitude of $\approx 0.2\%$, and hence the
maximum stellar jitter in the visible wavelengths should be no more
than $\approx 1$ m~s$^{-1}$.

Transiting planets with relatively long orbital periods ($P>30$ days)
detected by {\it K2} sometimes suffer from the problem of ``stale
ephemerides'' due to the small number of observed transits.  Indeed,
only two transits of \tar d were observed by {\it K2}, leading to a
large uncertainty in its orbital period.  Follow-up transit
observations are encouraged to enable accurate long-term predictions
of transit times.  \tar\ could be a good target for the upcoming
CHEOPS space mission \citep[CHaracterizing ExOPlanet
  Satellite;][]{2013EPJWC..4703005B}, which is specifically designed
to observe low-amplitude transits around bright stars.  The orbital
periods of \tar b and \tar c appear to be close to a 2:1 mean-motion
resonance and those of \tar c and \tar d are close to a 3:1 resonance,
but no clear signs of TTVs were seen in the $O-C$ diagrams
(Figure \ref{fig:ttv}).  It will be interesting to see if future
transit observations reveal any TTVs in this system.

\begin{figure}
\centering \includegraphics[width=8.5cm]{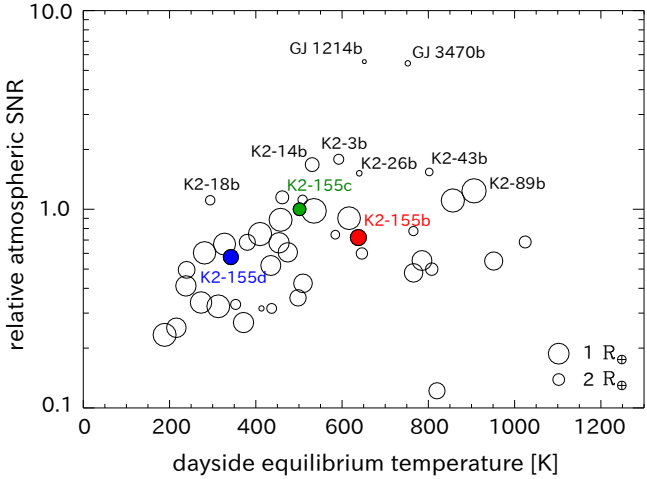}
\caption{
Relative SNR of transmission spectroscopy for known transiting planets ($R_p<6.0\,R_\oplus$)
around M dwarfs, 
calculated based on the stellar and planet radii, atmospheric scale height, $V$-band
magnitude, and transit duration.
The SNR's for \tar b, c, and d are plotted with the red, green, and blue circles,
respectively.
}
\label{fig:snr}
\end{figure}
The brightness of \tar\ also facilitates transit spectroscopy as a means of probing 
the atmospheres of the super-Earths. Following \citet{2017AJ....154..266N}, we plot in 
Figure \ref{fig:snr} the ``relative SNR" of transmission spectroscopy 
for known transiting planets except hot Jupiters ($R_p<6.0\,R_\oplus$) 
around M dwarfs, 
based on the stellar and planet radii, atmospheric scale height, $V$-band magnitude, and transit duration
\citep[see Equation (1a) and (1b) of][]{2017AJ....154..266N}.
We here plotted the SNR per transit rather than the SNR for a given
period of time as in \citet{2017AJ....154..266N}.
The three planets around \tar\ are plotted with the colored circles. 
Many Neptune-class planets ($R_p=2.0-6.0\,R_\oplus$; according to {\it Kepler}'s classification) 
show a higher SNR, but among super-Earths and 
Earth-like planets ($R_p<2.0\,R_\oplus$), \tar c is one of the best targets for
transmission spectroscopy. Figure \ref{fig:snr} also shows that \tar d is a good target 
in the sample as a possibly habitable super-Earth.

The actual signal amplitudes of the three super-Earths depend on the (unknown) scale heights of
their atmospheres. The atmospheric feature in transmission spectroscopy 
is of order $10H\cdot R_p/R_\star^2$, where $H$ is the atmospheric scale height
\citep{2009ApJ...690.1056M}. When a cloud-free hydrogen-dominated atmosphere is assumed,
the variation amplitude in transit depth
is expected to be $60$ ppm (\tar d) to $120$ ppm (\tar c),
which would be detectable by observations from the space (e.g., {\it Hubble} Space Telescope).
But if the planets have an Earth-like atmosphere (i.e., mean molecular weight of $\mu \sim 30$),
the expected signal would be $4-8$ ppm and its detection would be challenging.

\section{Summary\label{s:summary}}\label{s:summary}

In this paper, we have identified \tar, a relatively bright M dwarf
observed in the {\it K2} Campaign field 13, as a candidate planetary
system with three transiting super-Earths, and validated all these
planets based on speckle imaging and high resolution spectroscopy.
The coordinates of \tar\ are similar to that of the Hyades cluster,
but our spectroscopy indicates that its metallicity
($\mathrm{[Fe/H]}=-0.42\pm0.12$) is too low for a Hyades member, and
the RV and proper motions are also inconsistent with those of Hyades
members.  Indeed, \tar\ is one of the most metal-poor M-dwarf planet
hosts, which along with its long rotation period ($\approx 46$ days)
suggests that it is significantly older than the Hyades.

\tar d resides in or near the habitable zone, which led us to perform
3D global climate simulations to estimate the surface temperature of
\tar d assuming that the planet has an Earth-like composition and
atmosphere. 
We found that if the stellar insolation on \tar d is smaller than $1.5\,S_\oplus$, 
the planet could maintain a moderate climate with the averaged surface 
temperatures of $\lesssim 20\,^\circ$C and the stratospheric water vapor 
mixing ratio comparable to or below the classical moist greenhouse limit. 
The stellar insolation on \tar d has a large
uncertainty ($S_p=1.67\pm 0.38\,S_\oplus$) and thus its actual
habitability is not known at this point, but given the brightness of
the host star, this possibly habitable planet as well as the inner two
planets in this system are good targets for future follow-up studies
including Doppler mass measurements and transmission spectroscopy.

\acknowledgments 
We thank Adrian Price-Whelan for advice on stellar kinematics.
We thank the NOT staff members, in particular Peter S{\o}rensen, 
for their help, and support during the observations. 
Based on observations made with the Nordic Optical Telescope, 
operated by the Nordic Optical Telescope Scientific Association at the 
Observatorio del Roque de los Muchachos, La Palma, Spain, 
of the Instituto de Astrofisica de Canarias.
Data presented herein were obtained at the WIYN Observatory from
telescope time allocated to NN-EXPLORE through the scientific
partnership of the National Aeronautics and Space Administration, the
National Science Foundation, and the National Optical Astronomy
Observatory, obtained as part of an approved NOAO observing program
(P.I. Livingston, proposal ID 2017B-0334). NESSI was built at the Ames
Research Center by Steve B. Howell, Nic Scott, Elliott P. Horch, and
Emmett Quigley.  
This work was supported by Japan Society for
Promotion of Science (JSPS) KAKENHI Grant Number JP16K17660.
M.\,E. and W.\,D.\,C. were supported by NASA grant NNX16AJ11G to The University of Texas at Austin.
D.\,G. gratefully acknowledges the financial support of the
\emph{Programma Giovani Ricercatori -- Rita Levi Montalcini -- Rientro
  dei Cervelli (2012)} awarded by the Italian Ministry of Education,
Universities and Research (MIUR). 
This project has received funding from the European Union's Horizon 2020 
research and innovation programme under grant agreement No 730890. 
This material reflects only the authors views and the Commission is not liable 
for any use that may be made of the information contained therein.
I.R.\ acknowledges support by the Spanish Ministry of Economy and Competitiveness (MINECO) and the Fondo Europeo de Desarrollo Regional (FEDER) through grant ESP2016-80435-C2-1-R, as well as the support of the Generalitat de Catalunya/CERCA programme.
The authors are honored to be permitted to conduct observations on Iolkam Du'ag (Kitt Peak), a mountain within the Tohono O'odham Nation with particular significance to the Tohono O'odham people.

\software{IRAF \citep{1986SPIE..627..733T, 1993ASPC...52..173T}, 
SpecMatch-Emp \citep{2017ApJ...836...77Y},
PHOENIX \citep{2013MSAIS..24..128A}, 
vespa \citep{2012ApJ...761....6M, 2015ascl.soft03011M}
ROCKE-3D \citep{2017ApJS..231...12W}
}




\end{document}